\documentclass[twocolumn,superscriptaddress,preprintnumbers,amsmath,amssymb,prb]{revtex4}
\usepackage{graphicx}
\usepackage{dcolumn}
\usepackage{bm}
\usepackage[utf8]{inputenc}
\usepackage[T1]{fontenc}
\usepackage{mathptmx}
\usepackage{etoolbox}
\usepackage{amsmath}\usepackage{slashed}
\usepackage{color}
\usepackage{txfonts}

\begin{document}
\newcommand{\tm}{}
\preprint{AIP/123-QED}

\title{The Casimir free energy of peptide films on a silicon substrate:
Impact of dielectric-to-metal transition in silicon and nanoparticles in peptide}

\author{
G.~L.~Klimchitskaya}
\affiliation{Central Astronomical Observatory at Pulkovo of the Russian Academy of Sciences, St.Petersburg,
196140, Russia}
\affiliation{Peter the Great Saint Petersburg
Polytechnic University, Saint Petersburg, 195251, Russia}

\author{
V.~M.~Mostepanenko{$^*$}}
\affiliation{Central Astronomical Observatory at Pulkovo of the Russian Academy of Sciences, St.Petersburg,
196140, Russia}
\affiliation{Peter the Great Saint Petersburg
Polytechnic University, Saint Petersburg, 195251, Russia}
\email{vmostepa@gmail.com}

\date{\today}

\begin{abstract}
Using the Lifshitz theory of the van der Waals and Casimir forces, we calculate the Casimir  free
energy of thin peptide films deposited on silicon substrates. The Casimir free energy is found
as a function of film thickness for different fractions of water in the film, in the presence of either nonmagnetic or magnetic nanoparticles,
and under the impact of irradiation of a silicon substrate with laser pulses
or dopants resulting in the dielectric-to-metal phase transition. It is shown that for
a dielectric silicon there is the borderline value of the film thickness, such that the Casimir free
energy is negative and contributes to the film stability for thicker films, but is positive and makes
the film less stable for thinner ones. According to our results, the borderline value of peptide
film thickness decreases with increasing volume fractions of water and nanoparticles in the film.
This decrease is more pronounced for the magnetic nanoparticles and becomes stronger with
increasing their radius. The borderline value of peptide film thickness is found as a function
of the fraction of water in the film. If the silicon substrate is in metallic state, the Casimir free
energy of peptide coating is always positive, which makes it less stable. Possible applications
of the obtained results in organic electronics and biomedicine are discussed.
\end{abstract}

\maketitle

\section{INTRODUCTION}
The Casimir free energy and force originate from the zero-point and thermal fluctuations
of the electromagnetic field (see the seminal paper by Casimir\cite{1} and recent
monographs\cite{2,3,4}). It is usual to investigate the Casimir force acting between two
closely spaced bodies (thick plates, for instance) separated by a narrow gap. This might be
a vacuum gap or a gap filled with some material. In such an event, there is a density of the
Casimir energy per unit area of the plates. The Casimir force per unit area is found as the
negative derivative of this energy density with respect to the gap width. If the gap is filled
with a material, but the boundary plates are replaced with vacuum semispaces, the nonzero
Casimir energy density still persists. This is the Casimir energy (or the free energy if the
temperature is not equal to zero) of a freestanding material film having the same thickness
as the gap width. In a similar way, if only one plate is replaced with the vacuum semispace,
we obtain the Casimir free energy per unit area of a film deposited on another plate, which
plays the role of a substrate.

Using the Lifshitz theory of the van der Waals and Casimir forces,\cite{5,6,7} the
fluctuation-induced energy and corresponding pressure in liquid films of a poorly
wetted substrates were considered in Ref.~\onlinecite{8}. The systematic investigation of
the Casimir free energy of both metallic and dielectric films, either freestanding in vacuum
or deposited on substrates, was performed in Refs.~\onlinecite{9,10,11,12,13,14}. It was
shown that the Casimir free energy of a film may have different signs, i.e., the
electromagnetic fluctuations may make the film more stable if the Casimir free energy is
negative (this contributes to an attraction to the substrate) or less stable if it is positive.

During the last decades, considerable attention was paid to the electronic microdevices
containing organic elements, i.e., peptides, proteins, and other
biopolymers.\cite{15,16,17,18,19,20,21,22} The same elements found prospective applications
in biomedicine.\cite{23,24,25} The major contribution to investigation of various quantum
effects, including the effect of electromagnetic fluctuations in peptides, proteins and other
biostructures, was made by Professor Dr. Rudolf Podgornik.\cite{26,27,28,29,30,31,32,33}

With decreasing thickness of organic films used in microelectronics and biomedicine to
below a micrometer, the Casimir free energy begins to play a role in the film stability.
Using the Lifshitz theory, the Casimir free energy of freestanding in vacuum and deposited
on dielectric (SiO$_2$) and metallic (Au) substrates peptide films was found in
Refs.~\onlinecite{34} and \onlinecite{35}.
According to the obtained results, the Casimir free energy of
a freestanding peptide film is negative and, with increasing fraction of water in the film,
the free energy increases in magnitude. Thus, the effect of electromagnetic fluctuations
makes the film more stable. For a peptide film deposited on a SiO$_2$ substrate, the
Casimir free energy is negative for sufficiently thick films (typically thicker than
100~nm), but becomes positive for thinner films. In the case of metallic
substrate, the free energy of a peptide coating turned out to be positive and, thus,
makes the film less stable. Similar results were obtained for peptide films deposited
on GaAs, Ge, and ZnS semiconductor substrates, which may be in either metallic or
dielectric states depending on the level of doping.\cite{36} However, the Casimir free
energy of peptide films deposited on the silicon substrates was not investigated so far.
It was shown also that if the peptide coating on the SiO$_2$ substrate is doped with
either nonmagnetic or magnetic nanoparticles, the effect of the electromagnetic
fluctuations makes the film more stable down to smaller film thicknesses.\cite{37}

During the last few years, it was found that peptides have a broad spectrum of
antimicrobial properties including against the antibiotic-resistant microorganisms.
In this connection, there is currently the wide abundance of silicon medical
instruments, urinary tract catheters, for instance, coated with thin peptide films
and related studies.\cite{38,41,42,43,45,46,47,48,49} As a consequence, there are
additional reasons to calculate the Casimir free energy of peptide films deposited on
a silicon substrate and investigate its impact on the film stability under various
conditions.

In this paper, we consider the peptide coatings containing different fractions of
water on thick substrates made of dielectric silicon. Using the Lifshitz theory and the
frequency-dependent dielectric permittivity of a typical peptide, the Casimir free
energy of peptide coating is computed as a function of its thickness. The borderline
thickness of peptide coating possessing the zeroth Casimir free energy is found as a
function of the fraction of water in the film. For thicker peptide coatings, the Casimir
free energy is negative and contributes to the film stability, but for thinner coatings
the Casimir free energy is positive and makes it less stable. Next, the impact of dopants
or irradiation of the silicon substrate with laser pulses transforming silicon into a
metallic state on the stability of peptide coating is considered. It is shown that in
the metallic state of a silicon substrate the Casimir free energy of peptide coating with
any thickness becomes positive by making the coating less stable. We have also
considered the silicon substrate coated with peptide films, which are doped by either
nonmagnetic or magnetic nanoparticles. It is shown that in both cases the Casimir
free energy contributes to the film stability for thinner peptide coatings.

The paper is organized as follows. Section II contains a brief summary of the formalism
of Lifshitz's theory in application to thin peptide films deposited on a silicon plate and
the computational results for the Casimir free energy of peptide coatings containing
different fractions of water. The case of a silicon substrate in metallic state is also
considered. In Sec. III, the Casimir free energy of peptide coatings doped with either
nonmagnetic or magnetic nanoparticles is calculated. In Sec. IV, the thickness of
peptide coating on a silicon plate, which ensures the zeroth Casimir free energy, is found
as a function of the fraction of water contained in the coating in connection with the
problem of stability of peptide-coated silicon surfaces. Section V contains our
conclusions and a discussion.

\section{The Casimir free energy of peptide coatings on silicon substrate and the effect
of dielectric-to-metal phase transition}

\newcommand{\xk}{(i\xi_l,k)}
\newcommand{\xl}{(i\xi_l)}
\newcommand{\ve}{\varepsilon}

We consider the silicon substrate (thick plate) coated with thin peptide film of thickness $a$.
For a plate thickness exceeding approximately $2~\muup$m, the plate can be replaced with a
silicon semispace, which leads to the same results for the Casimir free energy.\cite{3}
Thus, the peptide-coated silicon plate presents a three-layer system, where the first layer
is the vacuum (air) semispace, the second layer is the peptide film of thickness $a$ and
the third layer is the silicon semispace.

Under an assumption that the peptide-coated plate is in thermal equilibrium with the
environment at temperature $T$, the Lifshitz theory provides the following expression for
the Casimir free energy of peptide film:\cite{3,12}
\begin{equation}
{\cal F}_p(a)=\frac{k_BT}{2\pi}\sum_{l=0}^{\infty}{\vphantom{\sum}}^{\prime}
\int_{0}^{\,\infty}\!\!k\,dk
\nonumber
\end{equation}
\begin{eqnarray}
&&~~\times
\left\{\ln\left[1-r_{\rm TM}^{(2,1)}\xk r_{\rm TM}^{(2,3)}\xk\,
e^{-2ak^{(2)}\xk}\right]\right.
\nonumber \\
&&~~\phantom{\times}\left.
+\ln\left[1-r_{\rm TE}^{(2,1)}\xk r_{\rm TE}^{(2,3)}\xk\,
e^{-2ak^{(2)}\xk}\right]\right\}.
\label{eq1}
\end{eqnarray}
\noindent
Here, $k_B$ is the Boltzmann constant, $k$ is the wave vector projection on the plane
of peptide film, $\xi_l=4\pi^2 k_BTl/h$ with $l=0,\,1,\,2,\,\ldots$ are the Matsubara
frequencies, and the prime on the summation sign in $l$ divides by 2 the term with
$l=0$.

The reflection coefficients $r_{\rm TM,TE}^{(2,1)}$ describe the reflection of the
transverse magnetic ($p$-polarized) and transverse electric ($s$-polarized) electromagnetic
waves on the boundary surface between the peptide film and the vacuum (air).
In a similar way, the coefficients $r_{\rm TM,TE}^{(2,3)}$ describe the reflection of the
transverse magnetic and transverse electric waves on the boundary surface between the
peptide film and the silicon plate.

Explicitly, the reflection coefficients can be presented in the form\cite{3}
\begin{eqnarray}
&&
r_{\rm TM}^{(2,n)}\xk=\frac{\ve^{(n)}\xl k^{(2)}\xk-
\ve^{(2)}\xl k^{(n)}\xk}{\ve^{(n)}\xl k^{(2)}\xk+\ve^{(2)}\xl k^{(n)}\xk},
\nonumber \\
&&
r_{\rm TE}^{(2,n)}\xk=\frac{ k^{(2)}\xk-
\mu^{(2)}\xl k^{(n)}\xk}{k^{(2)}\xk+\mu^{(2)}\xl k^{(n)}\xk},
\label{eq2}
\end{eqnarray}
\noindent
where $n=1,\,3$, the dielectric permittivities are of the air, $\ve^{(1)}=1$, of peptide,
$\ve^{(2)}$, and of silicon, $\ve^{(3)}$. The magnetic permeability of peptide,
$\mu^{(2)}$, is not equal to unity only when the peptide film is doped with
magnetic nanoparticles. The quantity $k^{(n)}$ with $n=1,\,2$ and 3 entering
Eqs.~(\ref{eq1}) and (\ref{eq2}) is defined as
\begin{equation}
k^{(n)}\xk=\left[k^2+\ve^{(n)}\xl\,\mu^{(n)}\xl\,\frac{\xi_l^2}{c^2}\right]^{1/2},
\label{eq3}
\end{equation}
\noindent
where $\mu^{(1)}=\mu^{(3)}=1$.

For computations using Eqs.~(\ref{eq1})\,--\,(\ref{eq3}), it is necessary to know the
dielectric permittivities of peptide films containing various fractions of water
and of silicon over the wide ranges of pure imaginary frequencies from the microwave
to infrared and ultraviolet domains. For peptide films containing various fractions of
water this problem was discussed in Ref.~\onlinecite{34}. It was noted that there is
no sufficient information in the literature about dielectric properties of peptides
over the wide frequency regions (see Refs.~\onlinecite{29,50,51,52}).
It was suggested\cite{34} to calculate the Casimir free energy  for the film of the
electrically neutral 18-residue zinc finger peptide investigated by the methods of
molecular dynamics in Ref.~\onlinecite{50}.

In Ref.~\onlinecite{50} one can find information about the dielectric permittivity
of this peptide in the microwave domain. Specifically, it was found that
$\ve^{(2)}(0)=15.$ This information was supplemented\cite{34} by the findings of
Ref.~\onlinecite{29} coathored by Prof.\ Podgornik, where the imaginary part of the
frequency-dependent permittivity of the cyclic tripeptide RGD-4C was calculated in the
ultraviolet frequency region. Taking into account that the molecules of both
peptides are rather similar in size and shape, the dielectric data obtained for them
were combined. In so doing, the contribution of infrared frequencies was modeled
using the Ninham-Parsegian representation. The resulting dielectric permittivity
of the model peptide along the imaginary frequency axis, $\ve^{(2)}(i\xi)$,
used below is shown in Fig.~1 of Ref.~\onlinecite{34}.

For the dielectric permittivity of water along the imaginary frequency axis,
$\ve^{(w)}(i\xi)$, the representation of Ref.~\onlinecite{53} is used
(see Fig.~1 in Ref.~\onlinecite{34}). The dielectric permittivities of peptide
films containing the volume fraction $\Phi$ of water, $\ve_{\Phi}^{(2)}(i\xi)$,
was obtained \tm{in the approximation of effective homogeneous medium. It is
assumed that the molecules of peptide are distributed in water, which plays the
role of a plasticizer, and the dielectric permittivity of each of these
substances satisfies the Clausius-Mossotti equation\cite{53a}
\begin{equation}
\frac{\ve^{(w,2)}(i\xi)-1}{\ve^{(w,2)}(i\xi)+2}=
\frac{4\pi}{3}N^{(w,2)}\alpha^{(w,2)}(i\xi),
\label{eq3a}
\end{equation}
\noindent
where $N^{(w,2)}$ are the numbers of molecules of water and peptide per unit volume
and $\alpha^{(w,2)}$ are their polarizabilities. Then, the dielectric permittivity
$\ve_{\Phi}^{(2)}(i\xi)$ can be found from the following mixing} formula:\cite{54}
\begin{equation}
\frac{\ve_{\Phi}^{(2)}(i\xi)-1}{\ve_{\Phi}^{(2)}(i\xi)+2}=
\Phi\,\frac{\ve^{(w)}(i\xi)-1}{\ve^{(w)}(i\xi)+2}+
(1-\Phi)\,\frac{\ve^{(2)}(i\xi)-1}{\ve^{(2)}(i\xi)+2}.
\label{eq4}
\end{equation}
\noindent
The examples of the permittivity  $\ve_{\Phi}^{(2)}$ with $\Phi=0.1,\,0.25$ and
0.4 can be found in Fig.~2 of Ref.~\onlinecite{34}.

\tm{It is well to bear in mind, however, that Eq.~(\ref{eq3a}),
which is almost exact for gases, leads to some errors when applied to
liquids. We emphasize also that the approximation of a homogeneous medium
is sufficiently exact only in the long-wavelength limit, i.e., at relatively
low frequencies.\cite{53b,53c}
Because of this, Eq.~(\ref{eq4}) is in fact a tentative estimate and in practical
applications, which require high accuracy, the dielectric permittivity of
a peptide film containing some fraction of water should be found by
measuring its complex index of refraction over a sufficiently wide
frequency region.}

The dielectric permittivity of silicon along the imaginary frequency,
$\ve^{(3)}(i\xi)$, was obtained by means of the Kramers-Kronig relation from
the imaginary part of its permittivity along the real frequency axis expressed via
the tabulated optical data for the complex index of refraction of high-resistivity
silicon.\cite{55} The computations of the Casimir free energy of peptide films are
performed below at room temperature $T=300~$K. It has been known that at nonzero
temperature any dielectric material possesses some small but nonzero DC conductivity
which vanishes exponentially fast with decreasing $T$. It was shown, however, that
theoretical predictions of the Lifshitz theory are excluded by the measurement data
if the DC conductivity of dielectric materials is included in
computations.\cite{56,57,58,59,60} If the DC conductivity of dielectric material
is disregarded, the Lifshitz theory comes to an agreement with the measurement
data.\cite{56,57,58,59,60} What is more, calculations of the Casimir free energy
with included DC conductivity of dielectrics using the Lifshitz theory result in
violation of the Nernst heat theorem.\cite{14,61,62,63,64} For these reasons,
below we use the experimentally and theoretically consistent approach by
disregarding the DC conductivity of dielectric silicon when calculating the
free energy of peptide films.

\begin{figure}[b]
\vspace*{-1.5cm}
\hspace*{-1cm}
\includegraphics[width=5.0in]{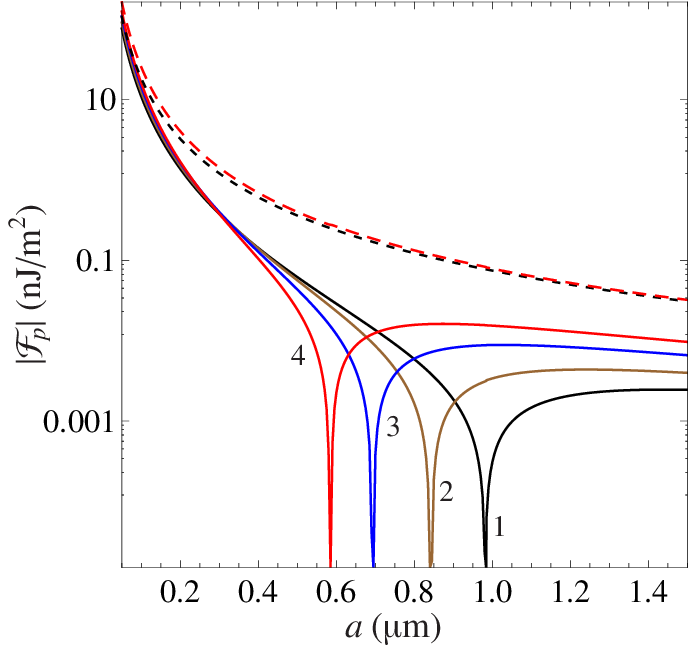}
\vspace*{-11.cm}
\caption{\label{fg1}
Magnitude of the Casimir free energy of peptide coatings containing
$\Phi = 0,~0.1,~0.25$, and 0.4 fractions of water deposited on a silicon plate in the dielectric
state is shown in the logarithmic scale as a function of peptide film thickness by the solid
lines 1, 2, 3, and 4, respectively. The lower and upper dashed lines show the Casimir free
energy of peptide films with $\Phi = 0$ and 0.4 fractions of water deposited on a silicon
plate in the metallic state, respectively.}
\end{figure}

The computational results for the magnitude of the Casimir free energy of peptide
films deposited on a silicon substrate are shown in the logarithmic scale in
Fig.~\ref{fg1} by the four solid lines as a function of film thickness. Line 1
is computed for the pure peptide film with no addition of water using the
dielectric permittivity $\ve^{(2)}(i\xi)$. For the borderline film thickness
$a_0=0.984~\muup$m the free energy of the film vanishes, it takes the
negative values at $a>a_0$ and the positive values for $a<a_0$. Thus,
the Casimir free energy of sufficiently thick peptide films contributes to
their stability, but for thinner films makes them less stable.

In the same figure, the solid lines 2, 3, and 4 show the Casimir free energy
for peptide coatings containing the fractions of water $\Phi=0.1$, 0.25, and
0.4, respectively. For the peptide films containing the indicated fractions
of water the Casimir free energy vanishes for the corresponding borderline
film thicknesses $a_0=0.835$, 0.695, and $0.583~\muup$m. For larger thicknesses,
the Casimir free energy is negative, for smaller --- positive. Thus, with increasing
fraction of water, the Casimir free energy contributes to the film stability  for
thinner films. From Fig.~\ref{fg1} it is also seen that with decreasing film thickness
down to $0.3~\muup$m an impact of the fraction of water on the Casimir free energy
becomes much smaller.

\begin{figure}[t]
\vspace*{-1.5cm}
\hspace*{-1cm}
\includegraphics[width=5.0in]{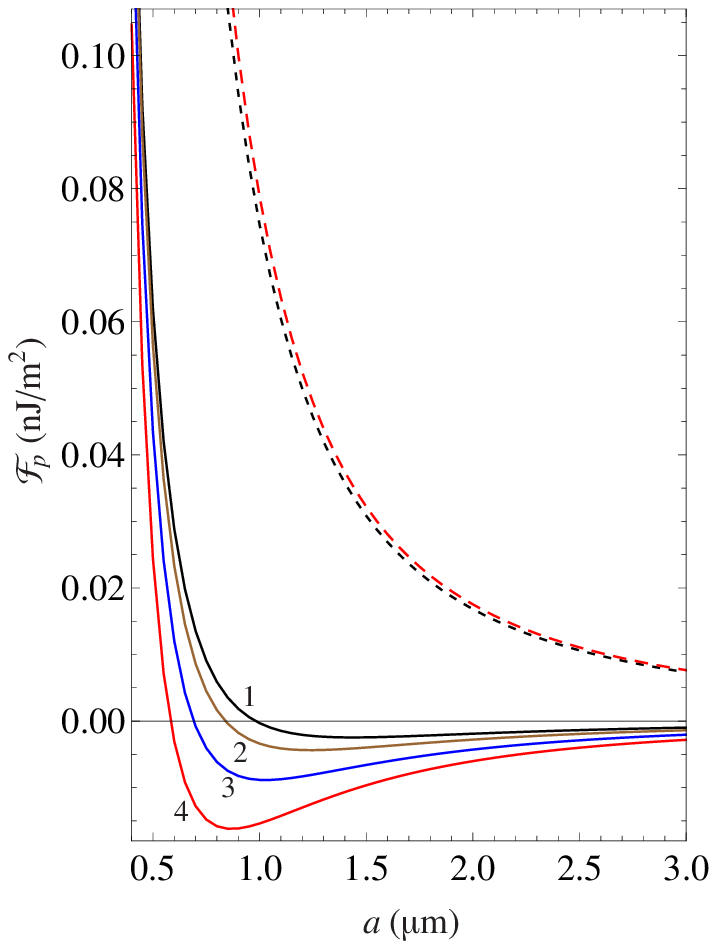}
\vspace*{-7.cm}
\caption{\label{fg2}
The Casimir free energy of peptide coatings containing
$\Phi = 0, 0.1$, 0.25, and 0.4 fractions of water deposited on a silicon plate in the dielectric
state is shown in the natural scale as a function of peptide film thickness by the solid lines
1, 2, 3, and 4, respectively. The lower and upper dashed lines show the Casimir free energy
of peptide films with $\Phi = 0$ and 0.4 fractions of water deposited on a silicon plate in
the metallic state, respectively.}
\end{figure}

A comparison with the previously obtained results for the Casimir free energy of
peptide films on a SiO$_2$ substrate\cite{34} shows that for a silicon substrate
it is much larger in magnitude. For instance, for a film of $a=100~$nm thickness with
$\Phi=0.25$ fraction of water the Casimir free energy of peptide film on a silicon
plate is by the
factor of 16 larger than the magnitude of the free energy on a  SiO$_2$ plate.
It should be also noted that the logarithmic scale used in Fig.~\ref{fg1} in
order to illustrate the case of sufficiently thin films down to $a=50~$nm disguises
the nonmonotonous behavior of the Casimir free energy for larger film thickness.
To illustrate it, in Fig.~\ref{fg2} we plot the Casimir free energy of peptide films
as a function of film thickness for larger film thicknesses with the same notations
as in Fig.~\ref{fg1}, but with its sign around the values, where the free energy
vanishes.

\tm{It is also interesting to compare the above results obtained for a silicon
substrate with the case of peptide-coated substrates made of some other semiconductor
materials.\cite{36} It should be taken into account, however, that
Ref.~\onlinecite{36} presents the computational results not in terms
of the Casimir free energy of peptide coating, but in terms of the Casimir
pressure in the film. To make a comparison, we consider the
minimum values on the lines 1, 2, 3, and 4 in Fig.~\ref{fg2} plotted for 0, 0.1,
0.25, and 0.4 fractions of water in the film, which are reached for the films
of 1.435, 1.24, 1.025, and 0.87 $\muup$m thickness, respectively. When
increasing the film thickness above these values, the Casimir pressure
changes its sign from positive to negative. To compare, for the ZnS
substrate the Casimir pressure changes its sign when the film thickness
becomes larger than 0.642, 0.598, 0.537, and 0.479 $\muup$m, respectively,\cite{36}
i.e., for significantly thinner peptide films. As one more example, for Ge
substrate the Casimir pressure in peptide coating with zero fraction of
water is positive (i.e., repulsive) for any film thickness and for the films
with 0.1, 0.25, and 0.4 fractions of water becomes negative for more than
3.359, 1.824, and 1.349 $\muup$m film thickness, respectively.\cite{36} Thus, for
different semiconductor substrates the impact of the Casimir effect on the
stability of peptide coating is quite different.}

Now we consider the case of a silicon substrate in the metallic state. This may be
reached either by using $p$ or $n$ dopants or by irradiation of the silicon substrate
with laser pulses.\cite{55a,55b,56,57,65} In the metallic state, the dielectric permittivity
of silicon along the imaginary frequency axis can be presented in the Drude-like form
\begin{equation}
\ve_m^{(3)}(i\xi)= \ve^{(3)}(i\xi)+\frac{4\pi\sigma(i\xi,T)}{\xi},
\label{eq5}
\end{equation}
\noindent
where the conductivity $\sigma$ is usually by several orders of magnitude larger than
the DC conductivity in the dielectric state and it does not vanish with vanishing
temperature.

{}From Eq.~(\ref{eq3}), it is easily seen that $k^{(3)}(0,k)=k$ irrespective of
whether the silicon plate is described by the dielectric permittivity $\ve^{(3)}$
or $\ve_m^{(3)}$. As a result, for a nonmagnetic coating $[\mu^{(2)}=1]$ one obtains
from Eq.~(\ref{eq2})
\begin{equation}
r_{\rm TE}^{(2,3)}(0,k)=r_{{\rm TE},m}^{(2,3)}(0,k)=0.
\label{eq6}
\end{equation}
\noindent
For the TM reflection coefficients, however, Eq.~(\ref{eq2}) leads to the following
different results for a silicon plate in the dielectric and metallic states:
\begin{equation}
r_{\rm TM}^{(2,3)}(0,k)=
\frac{\ve^{(3)}(0)-\ve_{\Phi}^{(2)}(0)}{\ve^{(3)}(0)+\ve_{\Phi}^{(2)}(0)},
\qquad
r_{{\rm TM},m}^{(2,3)}(0,k)=1.
\label{eq7}
\end{equation}

Computations show that the resulting difference in the zero-frequency contributions
to the Casimir free energy (\ref{eq1}) determines almost the total deviation between
the values of ${\cal F}_p$ and  ${\cal F}_{p,m}$ of the peptide films deposited on
the dielectric or metallic silicon substrates. Note that for two metallic test bodies
the Lifshitz theory using Eq.~(\ref{eq5}) is inconsistent with the measurement data
of precision experiments (see Refs.~\onlinecite{3}, \onlinecite{63}, and
\onlinecite{64} for a review). However, in
our case, when one of the plates is replaced with the air semispace, using the
plasma model extrapolation of the optical data to zero frequency leads to
approximately the same results.

In Figs.~\ref{fg1} and \ref{fg2}, the Casimir free energy of a pure peptide film
on a metallic silicon and a film containing $\Phi=0.4$ fraction of water are
shown as a function of film thickness by the lower and upper
dashed lines, respectively. The computational results for the films with $\Phi=0.1$
and 0.25 fractions of water are sandwiched between these two dashed lines.
The computations were performed as discussed above, but with the dielectric permittivity
of metallic silicon $\ve_m^{(3)}$ instead of $\ve^{(3)}$. In this case, the Casimir
free energy of peptide film remains positive for all the film thicknesses in the
entire region from 50~nm to $3~\muup$m, i.e., always makes the film less stable.

\section{THE CASIMIR FREE ENERGY OF PEPTIDE COATINGS DOPED WITH NANOPARTICLES}

Here, we consider the silicon substrates coated with peptide films doped with either
nonmagnetic or magnetic nanoparticles.  The procedure for fabrication the peptide films
with the homogeneously distributed metallic nanoparticles was elaborated in Ref.~\onlinecite{66}.
It was demonstrated\cite{37} that the presence of nanoparticles in peptide films deposited on
a SiO$_2$ substrate makes wider the range of film thicknesses, where the Casimir free energy
and pressure are negative. This leads to the increased film stability. Below we consider the same
effect for a silicon substrate.

Let the peptide film containing the fraction of water $\Phi$ is doped with spherical nanoparticles
of radius $R$ made of Au. It is assumed that nanoparticles occupy the volume fraction
$\varphi$ of the doped film. The dielectric permittivity of Au along the imaginary frequency axis,
$\varepsilon^{(\rm Au)}(i\xi)$, is obtained from its optical data for the complex index of
refraction\cite{55} and is often used in computations of the Casimir force.\cite{3,4,63,64,67,68}
The resulting dielectric permittivity of peptide film doped with nanoparticles can be obtained
from the Maxwell-Garnet mixing formula,\cite{53b,53c} which is designated for the dopants of
spherical shape
\begin{equation}
\varepsilon_{\Phi,\varphi}^{(2)}(i\xi) = \varepsilon_{\Phi}^{(2)}(i\xi)\left[1 + \frac{3\varphi X(i\xi)}
{1 - \varphi X(i\xi)}\right],
\label{eq8}
\end{equation}
\noindent
where
\begin{equation}
X(i\xi) = \frac{\varepsilon^{(\rm Au)}(i\xi) - \varepsilon_{\Phi}^{(2)}(i\xi)}
{\varepsilon^{(\rm Au)}(i\xi) + 2\varepsilon_{\Phi}^{(2)}(i\xi)}.
\label{eq9}
\end{equation}

\tm{There are also other mixing formulas of this type, for instance, the
Bruggeman mixing formula,\cite{53c} which is applicable to more than two
substances and is symmetric with respect to all the mixed components.
Application of the Bruggeman mixing formula to the case of three substances
(peptide films containing some fraction of water and doped with
nanoparticles) leads to slightly differing results as compared to
Eqs.~(\ref{eq8}) and (\ref{eq9}). However, as was
noted in Sec. II, in practical applications, which require high accuracy,
the dielectric permittivity of a peptide film should be found not by using
the mixing formulas, but by measuring its complex index of refraction over
a sufficiently wide frequency region.}

\begin{figure}[t]
\vspace*{-0.5cm}
\hspace*{-1cm}
\includegraphics[width=5.0in]{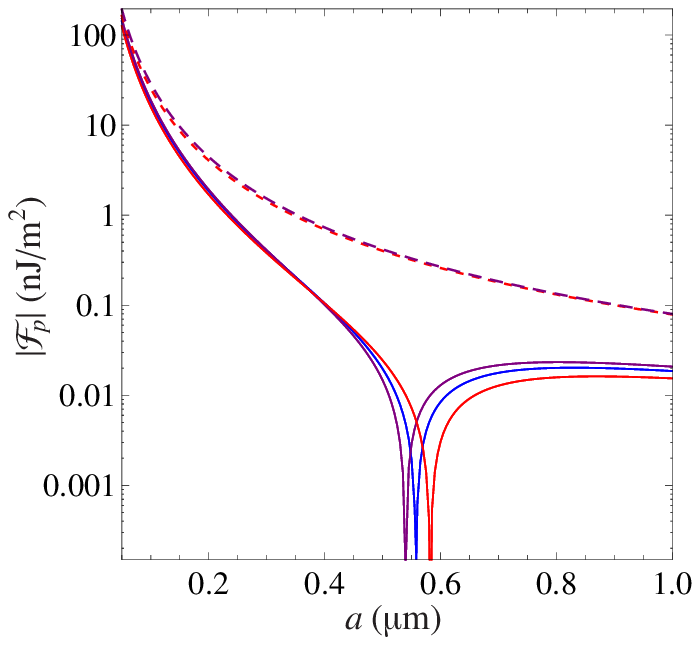}
\vspace*{-11.cm}
\caption{\label{fg3}
Magnitude of the Casimir free energy of peptide coatings containing
$\Phi = 0.4$  fraction of water and $\varphi = 0, 0.03$, and 0.05 fractions of Au nanoparticles
deposited on a silicon plate in the dielectric state is shown in the logarithmic scale as a
function of peptide film thickness by the three solid lines counted from right to left,
respectively. The two overlapping dashed lines show the Casimir free energy
of peptide films with $\Phi = 0.4$ fraction of water and $\varphi = 0$ and 0.05 fractions
of Au nanoparticles deposited on a silicon plate in the metallic state.}
\end{figure}

Computations of the Casimir free energy of peptide coating doped with Au nanoparticles on a
silicon substrate in the dielectric state is made in the same way as in Sec. II, but the dielectric
permittivity $\varepsilon_{\Phi}^{(2)}$ is replaced with $\varepsilon_{\Phi,\varphi}^{(2)}$. The
computational results for the magnitude of the Casimir free energy in the logarithmic scale
are presented in Fig.~\ref{fg3} as a function of separation by the three solid lines counted from
right to left for the peptide films containing $\Phi = 0.4$ volume fraction of water and
$\varphi = 0, 0.03$, and 0.05 fractions of Au nanoparticles, respectively. Thus, the right-most
solid line in Fig.~\ref{fg3} reproduces line 4 in Fig.~1 for comparison purposes.

As is seen in Fig.~\ref{fg3}, under the impact of nanoparticles, the borderline film thicknesses, where the
Casimir free energy of peptide coatings on a silicon plate vanishes, are shifted to smaller values.
Thus, for a film with $\varphi = 0.03$ fraction of nanoparticles (the medium solid line)
$a_0 = 0.558~\muup$m in place of $0.583~\muup$m as it was in the absence of nanoparticle
doping. For a film with $\varphi = 0.05$ fraction of nanoparticles (the left-most solid line),
 $a_0 = 0.535~\muup$m, i.e., with increasing fraction of nanoparticles in the film, $a_0$ decreases.
We recall that for $a > a_0$ the Casimir free energy takes negative values and contributes to
the film stability. This means that with decreasing $a_0$ a peptide film becomes more stable
over the wider range of film thicknesses. At the same time, for the film thicknesses
$a < 0.4~\muup$m and $a > 1~\muup$m an impact of nanoparticles on the Casimir free energy
gradually decreases. Note that the computational results for the Casimir free energy of
peptide coating doped with nonmagnetic nanoparticles do not depend on the nanoparticle
radius.

\tm{Note that for the case of a SiO$_2$ substrate\cite{37} the obtained numerical results
are quite different. Thus, the minimum values of the Casimir free energies, which
magnitudes are shown by the middle and left lines in Fig.~\ref{fg3}
plotted for the 0.03 and 0.05
fractions of Au nanoparticles in the films containing 0.4 fraction of water,
are reached for the films of 0.83 and 0.8 $\muup$m thickness, respectively.
The Casimir pressure in the film changes its sign from positive to negative
when the film thickness increases above these values. Under the same
conditions concerning the peptide coating, in the case of a SiO$_2$
substrate the Casimir pressure changes its sign when the film thickness
increases above 0.1 and 0.087 $\muup$m, respectively.\cite{37} This illustrates
the role of silicon as a substrate material.}

For a silicon substrate in the metallic state, the computational results for the Casimir free
energy of peptide coating with $\Phi = 0.4$ fraction of water are shown in Fig.~3 by the
two overlapping dashed lines plotted for $\varphi = 0$ (no doping) and $\varphi = 0.05$.
 It is seen that for a silicon substrate in the metallic state neither the presence
nor the value of doping in the peptide coating make a sizable impact on the Casimir free
energy of a film, which remains positive over the entire range of considered film thicknesses.

We are coming now to the case of peptide coatings doped with magnetic nanoparticles.
As an example, we consider the magnetite nanoparticles Fe$_3$O$_4$ frequently employed
in ferrofluids.\cite{70,71} Using the optical data of magnetite measured in Ref.~\onlinecite{72},
the dielectric permittivity of magnetite along the imaginary frequency axis,
$\varepsilon^{(\rm mg)}(i\xi)$, was found in Ref.~\onlinecite{71}. Then, using Eqs.~(\ref{eq8})
and (\ref{eq9}), and replacing there the dielectric permittivity of Au, $\varepsilon^{(\rm Au)}(i\xi)$,
with the dielectric permittivity of magnetite, $\varepsilon^{(\rm mg)}(i\xi)$, one obtains the
dielectric permittivity of a peptide film, $\varepsilon_{\Phi,\varphi}^{(2)}(i\xi)$, containing the
fraction of water $\Phi$ and doped with the fraction $\varphi$ of magnetic nanoparticles.

To calculate the Casimir free energy (\ref{eq1}) of doped peptide coatings in this case, one
also needs to know its magnetic permeability $\mu^{(2)}(i\xi)$. It was shown, however, that
the magnetic properties make an impact on the Casimir free energy and pressure only
through the contribution of the zero-frequency term in Eq.~(\ref{eq1}). This is because at
room temperature $\mu^{(2)}(i\xi)$ decreases with increasing $\xi$ and becomes equal to
unity already at much smaller $\xi$ than $\xi_1$.\cite{73}

The peptide coating, which contains the fraction $\varphi$ of single-domain magnetite
nanoparticles of radius $R$, is a superparamagnetic system. Its static magnetic
permeability is given by\cite{74}
\begin{equation}
\mu_{\varphi}^{(2)}(0) = 1 + \frac{16\pi^2 R^3\varphi M_s^2}{9k_BT},
\label{eq10}
\end{equation}
\noindent
where, according to Ref.~\onlinecite{75}, the saturation magnetization per unit volume of
a magnetite nanoparticle $M_s \approx 3\times 10^5$~A/m. Note that, according to
Eq.~ (\ref{eq10}), the magnetic permeability of peptide coating doped with magnetic
nanoparticles, unlike its dielectric permittivity  (\ref{eq8}), depends on the nanoparticle
radius $R$.

\begin{figure}[t]
\vspace*{-0.5cm}
\hspace*{-1cm}
\includegraphics[width=5.0in]{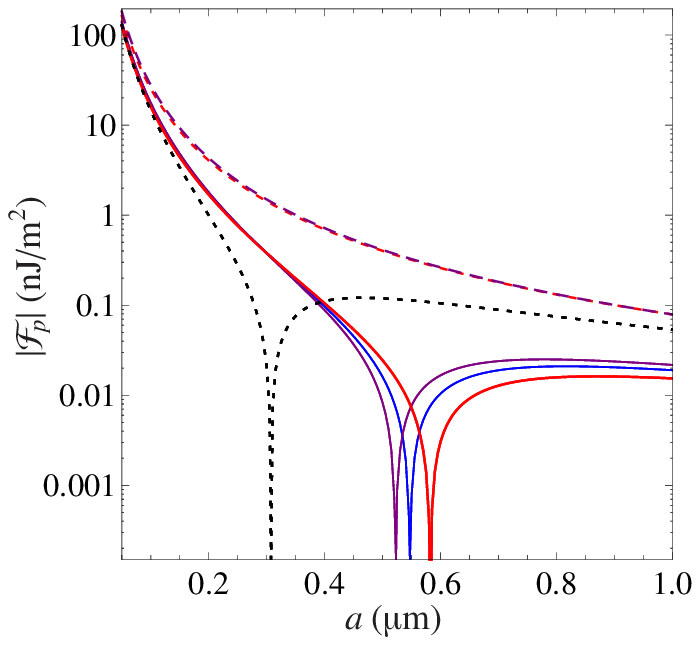}
\vspace*{-11.cm}
\caption{\label{fg4}
Magnitude of the Casimir free energy of peptide coatings containing
$\Phi = 0.4$  fraction of water and $\varphi = 0, 0.03$, and 0.05 fractions of magnetite
nanoparticles of 5~nm radius deposited on a silicon plate in the dielectric state is shown
in the logarithmic scale as a function of peptide film thickness by the three solid lines
counted from right to left, respectively. The extreme left dotted line shows the Casimir free
energy of peptide film with $\Phi = 0.4$  fraction of water and $\varphi = 0.05$ fraction
of magnetite nanoparticles of 7.5~nm radius deposited on a silicon plate in the dielectric
state. The two overlapping dashed lines show the Casimir free energy  of peptide films
with $\Phi = 0.4$ fraction of water and $\varphi = 0$ and 0.05 fractions of magnetite
nanoparticles deposited on a silicon plate in the metallic state.}
\end{figure}

Computations of the Casimir free energy of peptide coating doped with magnetic
nanoparticles on a silicon substrate in the dielectric state were performed by
Eqs.~ (\ref{eq1}) -- (\ref{eq3}) using the dielectric permittivity of peptide
$\varepsilon_{\Phi,\varphi}^{(2)}$ and its magnetic permeability $\mu_{\varphi}^{(2)}$.
The computational results for the magnitude of the Casimir free energy in the logarithmic
scale are presented in Fig.~\ref{fg4} as a function of separation by the three solid lines counted
from right to left for the peptide film containing $\Phi = 0.4$ fraction of water and
$\varphi = 0, 0.03$, and 0.05 fractions of magnetite nanoparticles of $R = 5$~nm radius,
respectively. The extreme left dotted line is plotted for the fraction $\Phi = 0.4$ of water
in the film and $\varphi = 0.05$ fraction of magnetite nanoparticles, but with larger
nanoparticle radius of $R = 7.5$~nm. The right-most solid line is again reproduced from
Fig.~\ref{fg1}, where it is shown as line 4.

According to Fig.~\ref{fg4}, the impact of magnetic nanoparticles further decreases the value of
$a_0$. Thus, if the peptide coating with $\Phi = 0.4$ fraction of water is doped with
$\varphi = 0.03$ fraction of magnetite nanoparticles of $R = 5$~nm radius, one obtains
$a_0 = 0.547~\muup$m in place of $a_0 = 0.558~\muup$m as it was in Fig.~\ref{fg3} in the case
of Au nanoparticles with the same radius and fraction. Under the same conditions, but
with $\varphi = 0.05$ fraction of magnetite nanoparticles, the borderline value of a
peptide film thickness is $a_0 = 0.523~\muup$m in place of $a_0 = 0.535~\muup$m as
for Au nanoparticles. Much larger changes in the value of $a_0$ are obtained for
magnetite nanoparticles with the same volume fraction $\varphi = 0.05$ but larger
radius. For example, for $R = 7.5$~nm we find from Fig.~\ref{fg4} $a_0 = 0.308~\muup$m (see the
extreme left dotted line). Thus, for a dielectric silicon substrate, the doping with magnetic
nanoparticles makes wider the region of peptide coating thicknesses, where the Casimir
free energy contributes to the film stability.

\tm{For a SiO$_2$ substrate coated with peptide films doped with magnetite nanoparticles,
the Casimir pressure in the film was found in Ref.~\onlinecite{37}. To compare with these
results, we note that the minimum values of the Casimir free energies, which magnitudes are
shown by the middle and left lines in Fig.~\ref{fg4} plotted for the 0.03 and 0.05 fractions
of magnetite nanoparticles with 5 nm radius in the film containing 0.4 fraction of water
are reached for the films of 0.815 and 0.78 $\muup$m thickness, respectively.
In the case of a SiO$_2$ substrate coated with the same peptide film,
the Casimir pressure changes its sign when the film thickness becomes larger
than 0.108 and 0.102 $\muup$m, respectively.\cite{37}}

In the case of a silicon substrate in metallic state coated with a peptide layer with $\Phi = 0.4$
fraction of water and doped with magnetic nanoparticles of 5~nm radius, the computational
results for the Casimir free energy as a function of film thickness are shown by two overlapping
dashed lines plotted for $\varphi = 0$ and  $\varphi = 0.05$. A comparison
with Fig.~\ref{fg3} allows to conclude that, if the silicon substrate is in a metallic state, the magnetic
properties of nanoparticles do not make a sizable impact on the value of the Casimir free
energy of peptide coating.

\section{BORDERLINE THICKNESS OF PEPTIDE COATING ON A SILICON SUBSTRATE}

According to the results obtained in Secs. II and III, for any fraction of water and either
magnetic or nonmagnetic nanoparticles contained in the peptide coating on a silicon
substrate in the dielectric state there is the borderline thickness $a_0$ such that the
Casimir free energy contributes to the coating stability for $a > a_0$ and makes it
less stable for $a < a_0$. Several examples of this kind are presented in
Figs.~\ref{fg1}--\,\ref{fg4}.
Here, we find the dependence of $a_0$ on the fraction of water contained in the
peptide coating under different assumptions about the character of doping.

\begin{figure}[b]
\vspace*{-3.5cm}
\hspace*{-1cm}
\includegraphics[width=5.0in]{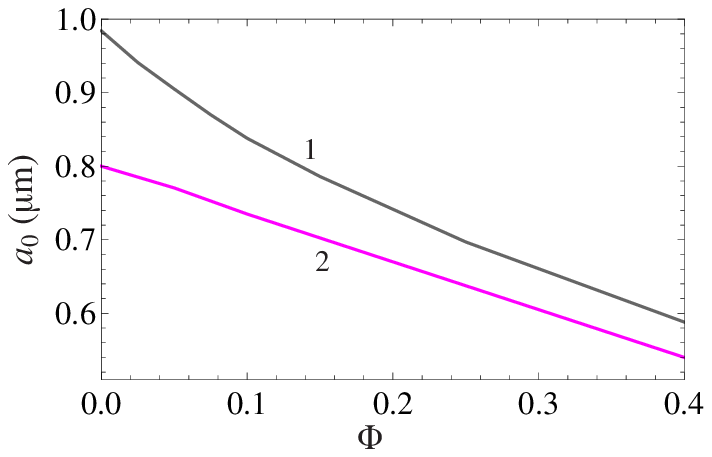}
\vspace*{-11.cm}
\caption{\label{fg5}
 The borderline thickness of peptide coating on a silicon plate in the dielectric
state, for which the Casimir free energy of peptide film vanishes, is shown as a function of
the fraction of water in the peptide film by the lines 1 and 2 for an undoped peptide and
doped with the 0.05 fraction of Au nanoparticles, respectively. In the region above each line,
the Casimir free energy contributes to the coating stability and below each line makes it less
stable.}
\end{figure}

Computations were again performed using Eqs.~(\ref{eq1})\,--\,(\ref{eq3}) with the
corresponding dielectric permittivity of the peptide coating. In each case, the Casimir free
energy ${\cal F}_p(a)$ was computed with a small step in the film thickness $a$ and the
value of $a = a_0$, where the free energy vanishes, ${\cal F}_p(a_0) = 0$, was determined.
We consider first an undoped peptide coating containing the fraction of water $\Phi$,
which is described by the dielectric permittivity $\varepsilon_{\Phi}^{(2)}$ (see Sec.~II).
The computational results for $a_0$ as a function of the fraction of water in the film are
shown by the line 1 in Fig.~\ref{fg5}. It is seen that $a_0$ monotonously decreases with
increasing $\Phi$, so that for a larger fraction of water in the peptide coating the
Casimir free energy contributes to the coating stability down to thinner films.

Next, we consider the peptide coating containing the fraction of water $\Phi$ and
doped with 0.05 fraction of Au nanoparticles. In this case, the coating is described by the
dielectric permittivity $\varepsilon_{\Phi,\varphi}^{(2)}$ (see Sec.~III). The computational
results for the borderline value of film thickness as a function of the fraction of water
is shown in Fig.~\ref{fg5} by the line 2. The borderline thickness $a_0$ again decreases with
increasing fraction of water, but for any $\Phi$ the region, where the Casimir free energy
 makes the film more stable, starts from thinner films.

\begin{figure}[h]
\vspace*{-3.5cm}
\hspace*{-1cm}
\includegraphics[width=5.0in]{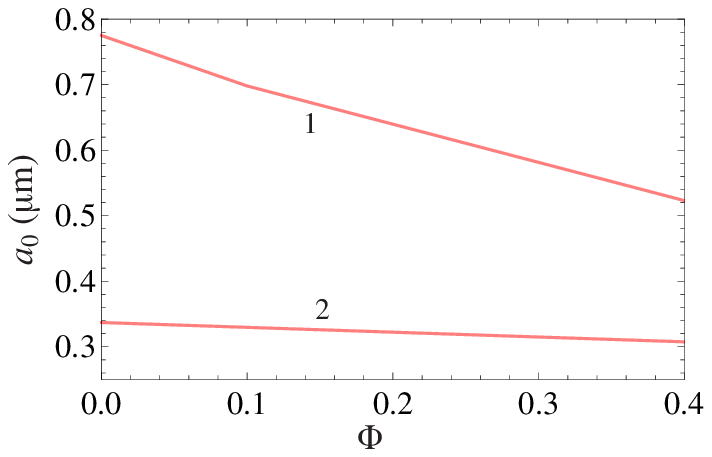}
\vspace*{-11.cm}
\caption{\label{fg6}
The borderline thickness of peptide coating on a silicon plate in the dielectric
state, for which the Casimir free energy of peptide film vanishes, is shown as a function of
the fraction of water in the peptide film by the lines 1 and 2 for a peptide coating
doped with the 0.05 fraction of magnetite nanoparticles having 5~nm and 7.5~nm radius,
respectively. In the region above each line, the Casimir free energy contributes to the
coating stability and below each line makes it less stable.}
\end{figure}

This effect is even more pronounced for the peptide films doped with magnetic
nanoparticles. For a $\varphi = 0.05$ fraction of magnetite nanoparticles in the peptide
coating, the dielectric permittivity $\varepsilon_{\Phi,\varphi}^{(2)}$  was obtained in
Sec.~III. In Fig.~\ref{fg6}, the computational results for $a_0$ as a function of the fraction
of water in peptide coating are shown by the line 1 for a nanoparticle radius
$R = 5$~nm and by the line 2 for $R = 7.5$~nm. The comparison of the line 1 in this
figure with the line 2 in Fig.~\ref{fg5} shows that the impact of magnetic properties of
nanoparticles decreases the borderline value of $a_0$ with fixed $\Phi$ only
slightly. The line 2 in Fig.~\ref{fg6}, however, shows that with increasing nanoparticle radius
for 50\% one obtains much smaller values of the borderline thickness of peptide
coating such that for thicker coatings the Casimir free energy contributes to the
coating stability.

From the practical point of view, it is important how much is the relative contribution
of the Casimir free energy to the total adhesion energy of peptide films deposited
on the silicon substrates. This energy contains several contributions determined by the
mechanical quality of both surfaces including the surface roughness, the chemical bonds,
electrical and diffusive interactions and, finally, the fluctuation-induced adhesion
caused by the van der Waals and Casimir forces.\cite{76,77,78,79} A crude estimation
shows\cite{8,34} that the Casimir free energy may contribute from 5\% to 20\% of the
total adhesion energy. Thus, the Casimir free energy of peptide films
deposited on the silicon substrates should be taken into account along with other
contributions to the cohesive energy when considering the problem of film stability.

\section{CONCLUSIONS AND DISCUSSION}

In the foregoing, we calculated the fluctuation-induced Casimir free energy of thin
peptide films deposited on a silicon substrate using the formalism of the Lifshitz
theory. This work is motivated by the expanding use of peptide-coated silicon
substrates in the organic electronics and biomedicine. In both these application
areas, it is desirable to have the sufficiently stable coatings.

According to our results, the Casimir free energy of peptide coatings on a silicon
substrate can be both negative and positive by making the peptide film more or
less stable, respectively, depending on the film thickness, the fraction of water in
the film, the rate of doping with nonmagnetic or magnetic nanoparticles, and
the dielectric properties of silicon. In the case of dielectric silicon, there is the
borderline thickness of peptide coating wherein the Casimir free energy vanishes.
It takes the positive values for thinner peptide films and the negative values for
thicker ones. With increasing volume fraction of water in the film, the value of the
borderline film thickness decreases.

We have also considered the peptide coatings on a silicon substrate doped with
both nonmagnetic and magnetic nanoparticles. Calculation of the Casimir free
energy of peptide film shows that, due to the presence of nanoparticles, the
electromagnetic fluctuations contribute to the film stability for thinner films.
The film becomes even more stable with increasing fraction of nanoparticles.
In doing so, the effect of magnetic nanoparticles is more pronounced than of
nonmagnetic ones. What is more, for the nonmagnetic nanoparticles, the decrease
in the borderline film thickness does not depend on a nanoparticle radius, whereas
for the magnetic ones the borderline thickness decreases significantly with
increasing nanoparticle radius. It should be noted also that some fraction of
magnetic nanoparticles makes the peptide coating on a silicon substrate
superparamagnetic, which may find applications in the organic electronics.

The above results were obtained for a high-resistivity silicon substrate in the
dielectric state. All computations of the Casimir free energy of peptide coatings
 were repeated when, under irradiation of the silicon substrate with laser pulses
or under adding some dopants, it undergoes a transition into the metallic state.
It is shown that in this case the Casimir free energy becomes positive, i.e., it
makes the peptide coating less stable for any film thickness and in the
presence of nanoparticles of any kind.

Finally, we obtained the dependence of the borderline thickness of peptide
coating on the fraction of water contained in the film. This was done for an
undoped peptide coating, as well as for a doped one with either nonmagnetic
or magnetic nanoparticles. It is shown that in all cases the borderline film
thickness imparting the zeroth Casimir free energy to the peptide coating is
the monotonously decreasing function of the fraction of water. With increased
radius of the magnetic nanoparticles, it becomes almost flat.

The obtained results allow evaluation of the role of Casimir free energy of thin
peptide films deposited on the silicon substrates under various conditions for
applications in organic electronics and biomedicine.

\begin{acknowledgments}
This work was supported by the State Assignment for Basic Research
(project FSEG-2026-0018).
\end{acknowledgments}


\begin{thebibliography}{99}
\bibitem{1}
H.~B.~G.~Casimir,
``On the attraction between two perfectly conducting plates,''
Proc. K. Ned. Acad. Wet. B {\bf 51}, 793 (1948).
\bibitem{2}
K.~A.~Milton,
{\it The Casimir Effect: Physical Manifestations of Zero-Point Energy}
(World Scientific, Singapore, 2001).
\bibitem{3}
M.~Bordag, G.~L.~Klimchitskaya, U.\ Mohideen, and
V.\ M.\ Mostepanenko,
{\it Advances in the Casimir Effect}
(Oxford University Press, Oxford, 2015).
\bibitem{4}
Bo E.~Sernelius,
{\it Fundamentals of van der Waals and Casimir Interactions}
(Springer, New York, 2018).
\bibitem{5}
E.~M.~Lifshitz,
``The theory of molecular attractive forces between solids,''
Zh. Eksp. Teor. Fiz. {\bf 29}, 94 (1955)
[Sov. Phys. JETP  {\bf 2}, 73 (1956)].
\bibitem{6}
I.~E.~Dzyaloshinskii, E.~M.~Lifshitz, and L.~P.~Pitaevskii,
``The general theory of van der Waals forces,''
Usp. Fiz. Nauk {\bf 73}, 381 (1961) [Adv. Phys. {\bf 10},165 (1961)].
\bibitem{7}
E.~M.~Lifshitz and L.~P.~Pitaevskii,
{\it Statistical Physics, Pt. II} (Pergamon Press, Oxford, 1984).
\bibitem{8}
I.~Boinovich and A.~Emelyanenko,
``Wetting and surface forces,''
Adv. Coll. Interface Sci. \textbf{165}, 60 (2011).
\bibitem{9}
G.~L.~Klimchitskaya and V.~M.~Mostepanenko,
``Casimir free energy of metallic films: Discriminating between Drude
and plasma model approaches,''
Phys. Rev. A \textbf{92}, 042109 (2015).
\bibitem{10}
G.~L.~Klimchitskaya and V.~M.~Mostepanenko,
``Casimir and van der Waals energy of anisotropic atomically thin metallic films,''
Phys. Rev. B \textbf{92}, 205410 (2015).
\bibitem{11}
G.~L.~Klimchitskaya and V.~M.~Mostepanenko,
``Casimir free energy and pressure for magnetic metal films,''
Phys. Rev. B \textbf{94}, 045404 (2016).
\bibitem{12}
G.~L.~Klimchitskaya and V.~M.~Mostepanenko,
``Characteristic properties of the Casimir free energy for metal films deposited on
metallic plates,''
Phys. Rev. A \textbf{93}, 042508 (2016).
\bibitem{13}
G.~L.~Klimchitskaya and V.~M.~Mostepanenko,
``Low-temperature behavior of the Casimir free energy and entropy of metallic films,''
Phys. Rev. A \textbf{95}, 012130 (2017).
\bibitem{14}
G.~L.~Klimchitskaya and V.~M.~Mostepanenko,
``Casimir free energy of dielectric films: Classical limit, low-temperature behavior and control,''
J. Phys.: Condens. Matter \textbf{29}, 275701 (2017).
\bibitem{15}
C.~D.~Dimitrakopoulos and P.~R.~L.~Malenfant,
``Organic thin film transistors for large area electronics,''
Advanced Materials \textbf{14}, 99 (2002).
\bibitem{16}
S.~Sharma, R.~W.~Johnson, and T.~A.~Desai,
``Evaluation of the stability of nonfouling ultrathin poly(elhylen glucol)
films for silicon-based microdevices,''
Langmuir {\bf 20}, 348 (2004).
\bibitem{17}
C.-Y.~Lee, J.-C.~Hwang, Y.-L.~Chueh, T.-H.~Chang, Y.-Y.~Cheng, and P.-C.~Lyu,
``Hydrated bovine serum albumin as the gate dielectric material for organic field-
effect transistors,''
Org. Electr. \textbf{14}, 2645 (2013).
\bibitem{18}
M.~Ma, X.~Xu, L.~Shi, and L.~Li,
``Organic field-effect transistors with a low driving voltage using albumin
as the dielectric layer,''
RSC Advances \textbf{4}, 58720 (2014).
\bibitem{19}
D.~T.~Simon, E.~O.~Gabrielsson, K.~Tybrandt, and M.~Berggren,
``Organic bioelectronics: Bringing the signaling between biology and technology,''
Chem. Rev. {\bf 116}, 13009 (2016).
\bibitem{20}
S.~S.~Panda, H.~E.~Katz, and J.~D.~Tovar,
``Solid-state electrical applications of protein and peptide based nanomaterials,''
Chem. Soc. Rev. {\bf 47}, 3640 (2018).
\bibitem{21}
P.~M.~Lee, Z.~Xiong, and J.~Ho,
``Methods for powering bioelectronic microdevices,''
Bioelectron. Med. (Lond.) {\bf 1}, 201 (2018).
\bibitem{22}
S.~A.~Moiz, M.~S.~Alshaikh, and A.~N.~M.~Alahmadi,
``Organic bioelectronics: Diversity of electronics along with biosensors,''
Biosensors {\bf 15}, 587 (2025).
\bibitem{23}
M.~Natesan and R.~G.~Ulrich,
``Protein microarrays, and biomarkers of infection disease,''
Int. J. Mol. Sci. \textbf{11}, 5165 (2010).
\bibitem{24}
A.~Sunna, A.~Care, and P.~L.~Bergquist,
{\it Peptide-Based Biomaterials and their Biomedical Apprications}
(Springer, Cham, 2018).
\bibitem{25}
Y.~Huang, K.~Yao, Q.~Zhang, X.~Huang, Z.~Chen, Y.~Zhou, and X.~Yu,
``Bioelectronics for electrical stimulation: materials, devices and biomedical applications,''
Chem. Soc. Rev. {\bf 53}, 8632 (2024).
\bibitem{26}
R.~Podgornik, H.~H.~Strey, K.~Gawrisch, D.~C.~Rau, A.~Rupprecht, and V.~A.~Parsegian,
``Bond orientational order, molecular motion, and free energy of high-density DNA mesophases,''
Proc. Nat. Acad. Sci. {\bf 93}, 4261 (1996).
\bibitem{27}
H.~H.~Strey, R.~Podgornik, D.~C.~Rau, and V.~A.~Parsegian,
``Dna-dna interactions,''
Curr. Opin. Struct. Biol. {\bf 8}, 309 (1998).
\bibitem{28}
L.~Javidpour, A.~L.~Bo{\u z}i{\u c}, A.~Naji, and R.~Podgornik,
``Multivalent ion effects on electrostatic stability of virus-like nano-shells,''
J. Chem. Phys. {\bf 139}, 154709 (2013).
\bibitem{29}
P.~Adhikari, A.~M.~Wen, R.~H.~French, V.~A.~Parsegian, N.~F.~Steinmetz, R.~Podgornik,
and W.-Y.~Ching,
``Electronic structure, dielectric response, and surface charge distribution of
 RGD (1FUV) peptide,''
Scient. Rep. \textbf{4}, 5605 (2014).
\bibitem{30}
B.-S.~Lu and R.~Podgornik,
``Effective interactions between fluid membranes,''
Phys. Rev. E \textbf{92}, 022112 (2015).
\bibitem{31}
A.~L.~Bo{\u z}i{\u c} and R.~Podgornik,
``pH dependence of charge multipole moments in proteins,''
Biophys. J. {\bf 113}, 1454 (2017).
\bibitem{32}
A.~L.~Bo{\u z}i{\u c} and R.~Podgornik,
``Site correlations, capacitance, and polarizability from protein protonation fluctuations,''
J. Phys. Chem. B {\bf 125}, 12902 (2021).
\bibitem{33}
W.~Broer, L.~Ge, F.~Xue, Y.~C.~Ren, and R.~Podgornik,
``Casimir-Lifshitz torque between cholesteric DNA stacks,''
Phys. Rev. B {\bf 113}, 085412 (2026).
\bibitem{34}
M.~A.~Baranov, G.~L.~Klimchitskaya, V.~M.~Mostepanenko, and E.~N.~Velichko,
``Fluctuation-induced free energy of thin peptide films,''
Phys. Rev. E {\bf 99}, 022410 (2019).
\bibitem{35}
V.~M.~Mostepanenko, E.~N.~Velichko, and M.~A.~Baranov,
``Role of electromagnetic fluctuations in organic electronics,''
J. Electr. Sci. Technol. {\bf 18}, 100023 (2020).
\bibitem{36}
G.~L.~Klimchitskaya, V.~M.~Mostepanenko, and O.~Yu.~Tsybin,
``Attractive and repulsive fluctuation-induced pressure in peptide films deposited on
semiconductor substrates,''
Symmetry {\bf 14}, 2196 (2022).
\bibitem{37}
G.~L.~Klimchitskaya, V.~M.~Mostepanenko, and E.~N.~Velichko,
``Effect of increased stability of peptide-based coatings in the Casimir regime via nanoparticle doping,''
Phys. Rev. B {\bf 102}, 161405(R) (2020).
\bibitem{38}
F.~Shamsi, H.~Coster, K.~A.~Jolliffe, and T.~Chilcott,
``Characterization of the substructure and properties of immobilized peptides on silicon surface,''
 Mater. Chem. Phys. {\bf 126}, 955 (2011).
\bibitem{41}
S.~K.~Ramakrishnan, M.~Martin, T.~Cloitre, L.~Firlej, and C.~Gergely,
``Molecular mechanism of selective binding of peptides to silicon surface,''
J. Chem. Inf. Mod. {\bf 54}, 2117 (2014).
\bibitem{42}
Z.~P{\'a}pa, S.~K.~Ramakrishnan, M.~Martin, T.~Cloitre, L.~Zim{\'a}nyi,
J.~M{\'a}rquez, J.~Budai, Z.~T{\'o}th, and T.~Gergely,
``Interactions at the peptide/silicon surfaces: Evidence of peptide multilayer,''
Langmuir {\bf 32}, 7250 (2016).
\bibitem{43}
S.~Franchi, C.~Battocchio, M.~Galluzzi, E.~Navisse, A.~Zamuner, M.~Dettin, and G.~Iucci,
``Self-assembling peptide hydrogels immobilized on silicon surfaces,''
Mater. Sci. Engin.: C {\bf 69}, 200 (2016).
\bibitem{45}
M.~N{\" o}th, Z.~Zou, I.~El-Awaad, L.~C. de Lencastre Novaes, G.~Dilarri,
M.~D.~Davari, H.~Ferreira, F.~Jakob, and U.~Schwaneberg,
``A peptide-based coating toolbox to enable click chemistry on polymers, metals,
and silicon through sortagging,''
Biotech. Bioeng. {\bf 118}, 1520 (2021).
\bibitem{46}
J.~Yan, P.~Siwakoti, S.~Shaw, S.~Bose, G.~Kokil, and T.~Kumeria,
``Porous silicon and silica carriers for delivery of peptide therapeutics,''
Drug Deliv. Translat. Res. {\bf 14}, 3549 (2024).
\bibitem{47}
I.~Soyhan, T.~Polat, E.~Mozioglu,  T.~A.~Ozal Ildeniz, M.~Acikel Elmas, S.~Cebeci, N.~Unubol, and O.~Gok,
``Effective immobilization of novel antimicrobial peptides via conjugation onto activated
silicon catheter surfaces,''
Pharmaceutics {\bf 16}, 1045 (2024).
\bibitem{48}
M.~Kosovari, T.~Buffeteau, L.~Thomas, A.-A.~Guay B{\' e}gin, L.~Vellutini,
J.~D.~McGettrick, G.~Laroche, and M.-C.~Durrieu,
``Silanization strategies for tailoring peptide functionalization on silicon surfaces:
Implications for enhancing stem cell adhesion,''
ACS Appl. Mater. Interf. {\bf 16}, 29770 (2024).
\bibitem{49}
K.~Awawdeh, X.~Jiang, L.~Dahan, M.~Atias, J.~Bahnemann, and E.~Segal,
``Porous silicon biosensors meet zwitterionic peptides: tackling biofouling from proteins to cells,''
Nanoscale Horiz. {\bf 10}, 3072 (2025).
\bibitem{50}
G.~L\"{o}ffler, H.~Schreiber, and O.~Steinhauser,
``Calculation of the dielectric properties of a protein and its solvent: Theory and a case study,
J. Mol. Biol. \textbf{270}, 520 (1997).
\bibitem{51}
T.~Sung, S.~D.~Namgung, J.~Lee, I.~R.~Choe, K.~T.~Nam, and J.-Y.~Kwon,
``Effects of proton conduction on dielectric properties of peptides,''
RSC Adv. {\bf 8}, 34047 (2018).
\bibitem{52}
J.~Dandurand, E.~Dantras, C.~Lacabanne, A.~Pepe, B.~Bochicchio, and V.~Samouillan,
``Thermal and dielectric fingerprints of self-assembling elastin peptides derived from exon30,''
AIMS Biophys. {\bf 8}, 236 (2021).
\bibitem{53}
L.~Bergstr\"{o}m,
``Hamaker constant of inorganic materials,''
Adv. Coll. Interface Sci. \textbf{70}, 125 (1997).
\tm{\bibitem{53a}
J.~D.~Jackson,
{\it Classical Electrodynamics}
(Wiley, New York, 1999).}
\bibitem{54}
D.~B.~Hough and L.~H.~White,
``The calculation of Hamaker constant from Lifshitz theory with
application to wetting phenomena,''
\bibitem{53b}
A.~H.~Sihvola,
{\it Electromagnetic Mixing Formulas and Applications}
(The Institution of Electrical Engineers, London, 1999).
\tm{\bibitem{53c}
V.~A.~Markel,
``Introduction to the Maxwell Garnett approximation: tutorial,''
J. Opt. Soc. Amer. A {\bf 33}, 1244 (2016).}
Adv. Coll. Interface Sci. \textbf{14}, 3 (1980).
\bibitem{55}
{\it Handbook of Optical Constants of Solids},
edited by E.~D.~Palik (Academic, New York, 1985).
\bibitem{55a}
J.~Opsal, M.~W.~Taylor, W.~L.~Smith, and A.~Rosencwaig,
``Temporal behavior of modulated optical reflection in silicon,''
J. Appl. Phys. {\bf 61}, 240 (1987).
\bibitem{55b}
T.~Vogel, G.~Dodel, E.~Holzhauer, H.~Salzmann, and A.~Theuer,
``High-speed switching of far-infrared radiation by photoionization in a semiconductor,''
Appl. Opt. {\bf 31}, 329 (1992).
\bibitem{56}
F.~Chen, G.~L.~Klimchitskaya, V.~M.~Mostepanenko, and U.~Mohideen,
``Demonstration of optically modulated dispersion forces,''
Optics Express {\bf 15}, 4823 (2007).
\bibitem{57}
F.~Chen, G.~L.~Klimchitskaya, V.~M.~Mostepanenko, and U.~Mohideen,
``Control of the Casimir force by the modification of dielectric properties with light,''
Phys. Rev. B {\bf 76}, 035338 (2007).
\bibitem{58}
G.~L.~Klimchitskaya and V.~M.~Mostepanenko,
``Conductivity of dielectric and thermal atom-wall interaction,''
J. Phys. A: Math. Theor. {\bf 41}, 312002 (2008).
\bibitem{59}
C.-C.~Chang, A.~A.~Banishev, G.~L.~Klimchitskaya, V.~M.~Mostepanenko, and U.~Mohideen,
``Reduction of the Casimir Force from Indium Tin Oxide Film by UV Treatment,''
Phys. Rev. Lett. {\bf 107}, 090403 (2011).
\bibitem{60}
A.~A.~Banishev, C.-C.~Chang, R.~Castillo-Garza, G.~L.~Klimchitskaya, V.~M.~Mostepanenko,
and U.~Mohideen,
``Modifying the Casimir force between indium tin oxide film and Au sphere,''
Phys. Rev. B {\bf 85}, 045436 (2012).
\bibitem{61}
B.~Geyer,  G.~L.~Klimchitskaya, and V.~M.~Mostepanenko,
``Thermal quantum field theory and the Casimir interaction between dielectrics,''
Phys. Rev. D {\bf 72}, 085009 (2005).
\bibitem{62}
G.~L.~Klimchitskaya and C.~C.~Korikov,
``Casimir entropy for magnetodielectrics,''
J. Phys.: Condens. Matter {\bf 27}, 214007 (2015).
\bibitem{63}
G.~L.~Klimchitskaya, U. Mohideen, and V.\ M.\ Mostepanenko,
``The Casimir force between real materials: Experiment and theory,''
Rev. Mod. Phys. {\bf 81}, 1827 (2009).
\bibitem{64}
V.\ M.\ Mostepanenko,
``Casimir Puzzle and Casimir Conundrum: Discovery and Search for Resolution,''
Universe {\bf 7}, 84 (2021).
\bibitem{65}
G.~L.~Klimchitskaya, U.~Mohideen, and V.~M.~Mostepanenko,
``Pulsating Casimir force,''
J. Phys. A: Math. Theor. {\bf 40}, 841 (2007).
\bibitem{66}
J.~Yan, Y.~Pan, A.~G.~Cheetham, Y.-A.~Lin, W.~Wang, H.~Cui, and C.-J.~Liu,
``One-step fabrication of self-assemled peptide thin films with highly dispersed
noble metal nanoparticles,''
Langmuir {\bf 29}, 16051 (2013).
\bibitem{67}
L.~M.~Woods, D.~A.~R.~Dalvit, A.~Tkatchenko, P.~Rodriguez-Lopez, A.~W.~Rodriguez, and R.~Podgornik,
``Materials perspective on Casimir and van der Waals interactions,''
Rev. Mod. Phys. {\bf 88}, 045003 (2016).
\bibitem{68}
R.~Esquivel-Sirvent, A.~Gusso, S.~Castillo-L{\' o}pez, and F.~S{\' a}nchez-Ochoa,
``Casimir force from Au to time crystals,''
Riv. Nuovo Cimento {\bf 49}, 137 (2026).
\bibitem{70}
T.~Guo, X.~Bian, and C.~Yang,
``A new method to prepare water based Fe$_3$O$_4$ ferrofluid with high stabilization,''
Physica A: Stat. Mech. Applic. {\bf 438}, 560 (2015).
\bibitem{71}
G.~L.~Klimchitskaya, V.~M.~Mostepanenko, E.~K.~Nepomnyashchaya,
and E.~N.~Velichko,
``Impact of magnetic nanoparticles on the Casimir pressure in three-layer systems,''
Phys. Rev. B {\bf 99}, 045433 (2019).
\bibitem{72}
A.~Schlegel, S.~F.~Alvarado, and P.~Wachter,
``Optical properties of magnetite (Fe$_3$O$_4$),''
J. Phys. C: Solid State Phys. {\bf 12}, 1157 (1979).
\bibitem{73}
B.~Geyer, G.~L.~Klimchitskaya, and V.~M.~Mostepanenko,
``Thermal Casimir interaction between two magnetodielectric plates,''
Phys. Rev. B {\bf 81}, 104101 (2010).
\bibitem{74}
S.~V.~Vonsovskii,
{\it Magnetism}
(Wiley, New York, 1974).
\bibitem{75}
S.~van Berkum, J.~T.~Dee, A.~P.~Philipse, and B.~E.~Ern{\' e},
``Frequency-dependent magnetic susceptibility of magnetite and cobalt
ferrite nanoparticles embedded in PAA hydrogel,''
Int. J. Mol. Sci. {\bf 14}, 10162 (2013).
\bibitem{76}
V.~E.~Basin,
``Advances in understanding the adhesion between solid substrates and organic coatings,''
Progr. Org. Coat. {\bf 12}, 213 (1984).
\bibitem{77}
B.~N.~J.~Persson and M.~Scaraggi,
``Theory of adhesion: Role of surface roughness,''
J. Chem. Phys. {\bf 141}, 124701 (2014).
\bibitem{78}
C.~Chen, S.~Bian, Y.~Jiang, L.~Yu, and J.~Hu,
``First principle calculations of interfacial adhesion energy between Fe substrate
and transition layer of films,''
Vacuum {\bf 218}, 112502 (2023).
\bibitem{79}
L.~Bricotte, K.~Chougrani, V.~Alard, V.~Ladmiral, and S.~Caillol,
``Adhesion theories: A didactic review about a century of progress,''
Int. J. Adhesion Adhesives {\bf 132}, 103673 (2024).

\end{thebibliography}
\end{document}